\documentclass[final,3p,times,twocolumn]{elsarticle}
\usepackage{amsmath}
\usepackage{graphicx}
\usepackage{amssymb}
\usepackage{gensymb}
\usepackage{natbib}
\usepackage{lineno}

\journal{Journal of Crystal Growth}

\begin{document}

\begin{frontmatter}
\tnotetext[fn1]{J. Cryst. Growth, accepted for publication, January 26, 2019}

\title{Double epitaxy of tetragonal and hexagonal phases in the FeSe system}

\author{S. B. Harris \corref{cor1}}
\ead{sumner@uab.edu}

\author{R. P. Camata \corref{cor2}}

\address{Department of Physics, University of Alabama at Birmingham, Birmingham, Alabama 35294, USA}
\cortext[cor1]{Corresponding author}

\begin{abstract}

Epitaxial films which contain more than one crystallographic phase or orientation are of interest due to the possibility of altered magnetic, electrical, and optical properties. Thin films of FeSe have been grown on single-crystal MgO substrates under conditions that produce the simultaneous, epitaxial growth of tetragonal and hexagonal phases. We show that this double epitaxy is characterized by phase domains with a well-defined epitaxial relationship to each other and that the relative phase fraction can be controlled. For growth temperatures of 350-450$\degree$C, the (001)-oriented tetragonal phase ($\beta$-FeSe) grows with its unit cell aligned with the cubic substrate, while a (101)-oriented hexagonal phase (Fe$_7$Se$_8$) shows domains with two different in-plane orientations separated by 45$\degree$. Additionally, the $\beta$-FeSe phase can be chosen to be (001)- or (101)-oriented with respect to the substrate, with the (101) orientation containing three rotated domains.
\end{abstract}


\end{frontmatter}


\section{Introduction}
\label{S:1}

Thin film heteroepitaxial growth on substrates of dissimilar lattice constant is an important area of crystal growth research due to applications-based restrictions on substrate choice and opportunities for tuning thin film properties via strain engineering. In heteroepitaxy, films generally release strain through a variety of mechanisms such as the formation of domains, dislocations, and mosaic structures. For some systems, a single crystalline phase may exhibit growth of domains with two different epitaxial relationships to the substrate.
This behavior, referred to as double epitaxy, has been observed in important systems such as the perovskites SrTiO$_3$ \cite{Tang2018} and LaFeO$_3$ \cite{Bi2009}, magnetic materials such as CoFe$_2$O$_4$ \cite{Chen2015} and superconductors such as YBCO \cite{Schlom1996}.
The presence of multiple epitaxial orientations creates materials with many interfaces at fixed angles with respect to the substrate and is known to alter thermal, electrical, and piezoelectric properties \cite{Macmanus-driscoll2010}.
A further development in double epitaxy is the ability to grow two distinct materials whose interspersed crystal domains exhibit well-defined epitaxial relationships to each other and to the substrate. This may allow controlled out-of-plane interfaces that enable interface-enhanced effects, microstructural configurations that stabilize non-equilibrium phases, and preservation of strain in thick, vertically aligned nanocomposite films \cite{QuanxiSciAdv}. This work demonstrates the control of double epitaxy in FeSe, a significant compound in the current efforts to elucidate superconductivity in iron-based superconductors. We show that in pulsed laser deposition (PLD) of FeSe on (001)-oriented MgO, conditions exist that lead to the epitaxial growth of tetragonal and hexagonal phases of FeSe in juxtaposed domains throughout the film. We investigate the impact of PLD parameters and chemical doping on this double epitaxy and identify the conditions under which it is suppressed.

At atmospheric pressure, FeSe crystallizes into numerous distinct phases. A stoichiometrically sensitive phase known as $\beta$-FeSe has a tetragonal unit cell and exhibits the anti-PbO-type structure (space group $P4/nmm$) \cite{Okamoto, McQueen2009ex}. Several hexagonal phases lie in close proximity to the $\beta$ phase in the Fe-Se phase diagram. These include the high-temperature stoichiometric $\delta$-FeSe phase, which has the NiAs structure (space group $P6_3mc$) \cite{Okamoto} and the lower-temperature Fe$_{7}$Se$_{8}$, which also has the fundamental NiAs structure, but in which one eighth of the Fe sites are vacant. The Fe vacancies in this nonstoichiometric phase may take on different ordered arrangements giving rise to two main types of Fe$_{7}$Se$_{8}$ superstructures that can be characterized by their unit cell length along the $c$-axis. One of the superstructures has a unit cell that is three times that of the fundamental NiAs-type cell (3c structure), while the other is four times (4c structure) \cite{Kawaminami1970}. Another hexagonal NiAs-type phase ($\alpha$-FeSe) has also been identified below approximately 300$\degree$C with larger lattice parameters than those for Fe$_{7}$Se$_{8}$ \cite{McQueen2009ex}.\par
The tetragonal $\beta$-FeSe phase has drawn much attention recently because of its superconducting properties with the highest observed critical temperature among iron-based superconductors ($T_c$ $\sim$ 109 K for $\beta$-FeSe monolayers on single-crystal SrTiO$_3$) \cite{Zhang2015}. Epitaxial (001)-oriented $\beta$-FeSe grown by PLD has been demonstrated on many different substrates and across a broad range of deposition conditions \cite{Feng2018}. Reports of epitaxial $\beta$-FeSe films with thickness greater than 200 nm often indicate the presence of secondary phases of FeSe. However, the effect of these secondary phases on the crystallization behavior of the predominantly $\beta$-FeSe films has not yet been systematically discussed. In this paper we explore these secondary-phase effects and identify the particular conditions under which tetragonal and hexagonal phases of FeSe grow with a well defined epitaxial relationship to each other and to the cubic MgO substrate.

\section{Experimental Details}
\label{S:2}

Thin films of FeSe were deposited on commercially available, single-side polished (001)-oriented MgO substrates (Sigma-Aldrich) by PLD using the focused beam of a KrF excimer laser (Lambda Physik LPX 305i). The symmetry and lattice parameters of the FeSe phases are conventionally unsuitable for lattice-matching epitaxial growth on (001) MgO. In the case of the $\beta$ phase for example, its lattice parameter ($a$ = 3.672-3.769 \AA, depending on exact stoichiometry) differs from that of MgO ($a$ = 4.211 \AA) by 11-13\%, which is greater than the maximum $\sim$ 9\% normally tolerated for epitaxy \cite{Narayan2003, Deng2018}. However, the cubic structure of MgO exhibits various domain-matching epitaxial relationships with the FeSe phases, opening opportunities for controlling epilayer orientation. In addition to this potential for orientation control, MgO substrates also offer excellent compatibility with planned spectroscopic studies of doubly epitaxial FeSe. The MgO substrates were prepared by boiling in acetone for 2 minutes followed by successive cleaning with acetone and methanol in an ultrasonic bath for 5 minutes each, followed by a N$_2$ blow dry, and promptly placed into the high-vacuum PLD chamber. This method of substrate preparation removes any undesirable Mg(OH)$_2$ formed due to ambient exposure \cite{LeFebvrier2017} and also eliminates other surface contaminants including hydrocarbons. Thin films were deposited under high vacuum (10$^{-7}$ torr) with various laser energy densities in the 1.4-3.4 J/cm$^2$ range obtained by focusing the 248-nm laser pulses to a spot with area of 0.13 cm$^2$. All films were deposited using a laser repetition rate of 10 Hz and a target-to-substrate distance of 55 mm. The substrate temperature of the individual samples was chosen between 350$\degree$C and 550$\degree$C. The PLD targets were prepared by pressing FeSe powder (99.99\%, American Elements) into 19 mm diameter discs at 3000 psi. The discs were then placed in quartz ampoules, evacuated to approximately 10 mtorr, sealed with a blowtorch, and sintered at 700$\degree$C for 12 hours. For evaluating the effect of chemical doping during epitaxy, copper (Cu) was chosen as a dopant and added to other PLD targets in various concentrations. In this case, each ``Cu-doped'' target was produced with a specific atomic percent of Cu with respect to Fe in the range between zero and 1.5 at.\% Cu by mixing the proper relative masses of FeSe and Cu$_2$Se (99.999\%, American Elements) precursor powders. This approach produces targets with lower Se content than the undoped targets. For an intended Cu atomic fraction $x$, FeSe and Cu$_2$Se are mixed in the ($1-x$):($x/2$) ratio, leading to a target composition of Fe$_{1-x}$Cu$_x$Se$_{1-x/2}$. The lower Se content may favor the growth of Cu-doped FeSe in the $\beta$ phase, since the corresponding undoped phase ($\beta$-FeSe) exists in the Fe-rich side of FeSe stoichiometry. Film thickness was targeted at 400 nm for all films, which is sufficient to allow for the appearance of secondary FeSe phases, and was determined by separate deposition rate calibration experiments using cross-sectional scanning electron microscopy for thickness measurements. Prior to deposition, the rotating targets were pre-ablated with 1000 pulses at 10 Hz to remove surface contamination, ensuring that the MgO substrates were shielded from ablated species generated during pre-ablation. Once the deposition was complete, the substrate heater was immediately turned off and left to cool to room temperature under vacuum. All films had a dark gray, highly reflective appearance typical of metallic films.
The films were characterized by X-ray diffraction (XRD) (PANalytical Empyrean, Cu K$\alpha$). The surface of each film was also studied using an atomic force microscope (AFM) operated in non-contact mode (NanoInk DPN 5000, tip radius \textless\ 10 nm).

\section{Results and Discussion}
\label{S:3}

\begin{figure}
    \includegraphics[width=7.8cm]{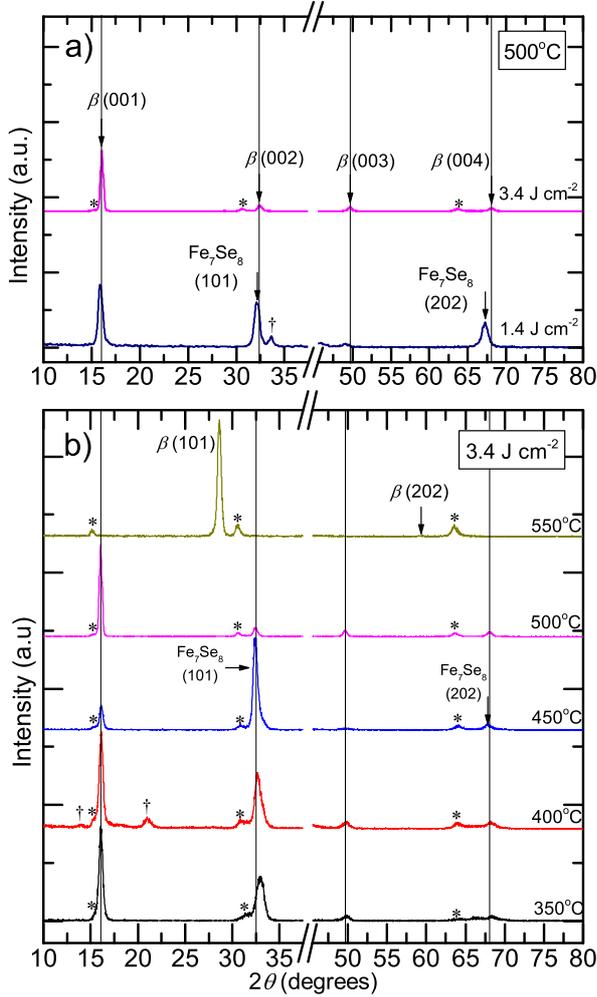}
    \caption{a) $\theta$-2$\theta$ scans of FeSe films grown with two different values of laser fluence at the same substrate temperature of 500$\degree$C. The film grown at high fluence (3.4 J/cm$^2$) is predominantly (001)-oriented $\beta$-FeSe with the minor presence of another epitaxially-oriented phase, as evidenced by the low-intensity reflections marked with asterisks ($\ast$). This minor phase is identified as hexagonal Fe$_7$Se$_8$. The film grown at low fluence (1.4 J/cm$^2$) is mixed phase (001)-oriented $\beta$-FeSe and Fe$_7$Se$_8$ with the [101] direction of its fundamental NiAs-type structure aligned normal to the substrate. b) $\theta$-2$\theta$ scans of FeSe films grown with a laser fluence of 3.4 J/cm$^2$ at different substrate temperatures show how the crystallographic phase preference changes with temperature. Films grown between 350$\degree$C and 450$\degree$C comprise (001)-oriented $\beta$-FeSe and (101)-oriented Fe$_7$Se$_8$ with relative fractions changing with temperature. As corroborated by subsequent $\phi$ scans, these two phases exist in a doubly epitaxial configuration. Films grown at 500$\degree$C and above are dominated by $\beta$-FeSe without the presence of the (101)-oriented Fe$_7$Se$_8$ phase. For deposition at 550$\degree$C the orientation of $\beta$-FeSe is noted to change from the typical $\beta$-FeSe (001)$\parallel$(001) MgO to $\beta$-FeSe (101)$\parallel$(001) MgO. Vertical lines represent the $\beta$-FeSe (00$\ell$) peak positions. The indexed Fe$_7$Se$_8$ reflections show Miller indices corresponding to the fundamental NiAs-type structure of Fe$_7$Se$_8$. Peaks marked with daggers ($\dagger$) are consistent with other Fe$_7$Se$_8$ reflections. The horizontal axis is broken to remove the very large (002) MgO substrate peak around 2$\theta$ = 43$\degree$.}
    \label{Fig 1.}
\end{figure}
Several FeSe films were grown to determine the effect of substrate temperature and laser energy density on the crystallographic phase and orientation of multi-phase, epitaxial FeSe on MgO. The FeSe films grown with laser energy density in the 1.4-3.4 J/cm$^2$ range and substrate temperatures between 350$\degree$C and 550$\degree$C exhibit a variety of crystallographic phases and orientations.

\begin{figure}[t]
    \includegraphics[width=7.8cm]{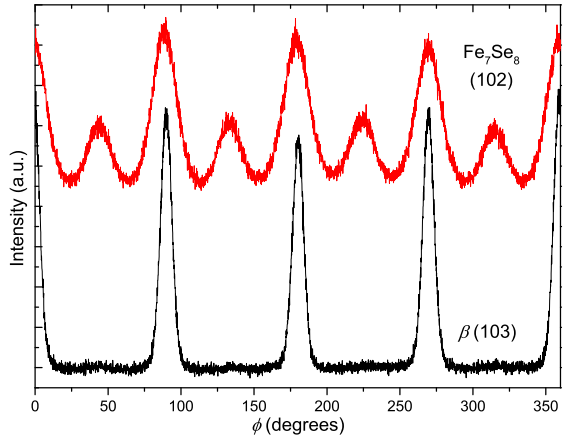}
    \caption{$\phi$ XRD scans of the asymmetric (103) reflection of $\beta$-FeSe and the (102) reflection of the fundamental NiAs-type structure of Fe$_7$Se$_8$. The four-fold symmetry of the scans indicates that the in-plane orientation of $\beta$-FeSe and Fe$_7$Se$_8$ are ``in line'' with the cubic MgO substrate, confirming domain-matching epitaxy for both phases. The Fe$_7$Se$_8$ phase also shows a minority domain which is rotated by 45$\degree$.}
    \label{fig:my_label}
\end{figure}

\begin{figure}
    \centering
    \includegraphics[width=7.8cm]{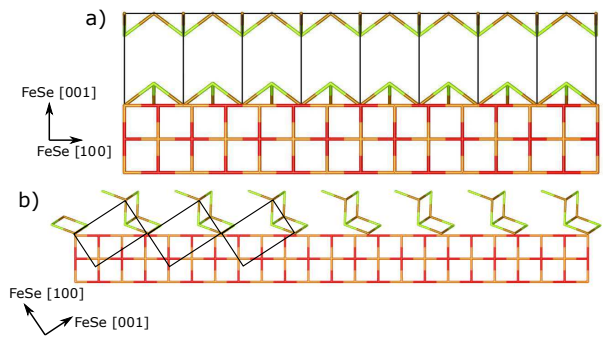}
    \caption{a) Schematic of the suggested 8/7 domain-matching epitaxial relationship for $c$-axis oriented $\beta$-FeSe on MgO. b) Possible arrangement for 7/11 domain-matching of (101)-oriented $\beta$-FeSe. The MgO substrate is the cubic structure with Mg and O on orange and red vertices, respectively. Fe and Se are represented by brown and green vertices, respectively. (Color online)}
    \label{Fig 2.}
\end{figure}

Figure \ref{Fig 1.}a shows $\theta$-2$\theta$ XRD scans for two thin films grown at a substrate temperature of 500$\degree$C with different values of laser energy density incident on the PLD target. The film grown with 3.4 J/cm$^2$ clearly produced the tetragonal $\beta$-FeSe phase with the (001) reflection around 2$\theta$ = 16$\degree$ accompanied by the (002), (003), and (004) reflections whose positions are indicated in the figure. This leads to the conclusion that this film is highly oriented with the epitaxial relation $\beta$-FeSe (001)$\parallel$(001) MgO. The peak position of the (001) family of planes is 16.064$\degree$ which corresponds to a $c$-axis of length 5.513 \AA, in agreement with literature values for unstrained bulk $\beta$-FeSe \cite{Margadonna2009}. Rocking curve analysis (Supplementary material) of the $\beta$-FeSe (001) reflection has FWHM of 1.3$\degree$, which is substantially broader than the diffractometer angular resolution of 0.08$\degree$ measured for single crystals at the same experimental conditions. This indicates mosaic structure of $\beta$-FeSe in the film, meaning individual grains are tilted with some small angle with respect to each other. The unstrained $c$-axis, mosaicity, as well as the epitaxial nature of this film are consistent with domain-matching epitaxial growth  \cite{Srikant1997}. A $\phi$ scan of the $\beta$-(103) asymmetric reflection shown in Fig. \ref{fig:my_label} reveals four-fold symmetry, in line with the (200) reflection of MgO (set to 0$\degree$). This indicates that the in-plane orientation is such that $\beta$-FeSe [100]$\parallel$[100] MgO, corroborating the evidence for $\beta$-FeSe epitaxy inferred from Fig. \ref{Fig 1.}a. This is true for all subsequent films reported in this paper that contain $\beta$-FeSe. Along the in-plane orientation of the $\beta$-FeSe phase, it is possible to conceive of domains with integer numbers of $\beta$-FeSe lattice planes that match similarly envisioned domains in (001)-oriented MgO. Many of these domain-matching relationships have mismatches substantially less than the 11-13\% obtained by comparing single unit cells. One likely scenario supported by the $\beta$-(103) $\phi$ scan, is $\beta$-FeSe [100]$\parallel$[100] MgO with 8 unit cells of FeSe matching to 7 unit cells of MgO as shown in Fig. \ref{Fig 2.}a. This domain matching relationship leads to an overall mismatch of 0.35\% using the lattice constants $a$ = 4.211 \AA\ for MgO and $a$ = 3.672 \AA\ for $\beta$-FeSe. 
This film shows the presence of another minor phase, which is also epitaxially oriented and is identified as hexagonal Fe$_7$Se$_8$.
This identification is based on the peak positions and corresponding interplanar distances of the three low-intensity reflections marked by asterisks ($\ast$) in Fig. \ref{Fig 1.}a. These reflections occur at 2${\theta}$ = 15.13$\degree$, 30.54$\degree$, and 63.56$\degree$, corresponding to interplanar distances of 5.850 $\AA$, 2.925 $\AA$, and 1.463 $\AA$, respectively. These distances are in the 4:2:1 ratio characteristic of parallel planes in a common crystal structure. They are consistent with the interplanar separations of the (00$\ell$) families of planes of a NiAs-type structure, for which every other plane of Fe atoms contains the vacant sites that characterize Fe$_7$Se$_8$ \cite{Kawaminami1970}. Such interplanar distances yield an equivalent, fundamental NiAs-type unit cell with $c$ = 5.850 $\AA$. The 3 reflections also indicate that the Fe$_7$Se$_8$ phase is oriented with the $c$-axis of its NiAs-type structure aligned with the direction normal to the surface. In the setting of the 3c-Fe$_7$Se$_8$ structure, these reflections would index to the (003), (006), and (00.12) families of planes, whereas for the 4c structure, the appropriate indexing for the same peaks would be (004), (008), and (00.16), respectively (Supplementary material).
The presence of hexagonal Fe$_7$Se$_8$ together with $\beta$-FeSe is consistent with the proximity of these two phases in the Fe-Se phase diagram and our PLD conditions. XRD measurements of our sintered FeSe targets revealed a mixture of $\beta$-FeSe ($\sim$ 22\%) and 3c-Fe$_{7}$Se$_{8}$ ($\sim$ 78\%), corresponding to a Se-enriched overall target composition (Supplementary material). It is therefore not surprising to find Fe$_7$Se$_8$ in the films. The dominant presence of $\beta$-FeSe reveals, however, that despite the use of a Se-enriched target, the characteristics of the PLD flux or the kinetics of the growth process allow for the Fe-rich $\beta$ phase to form in the sample grown at 500$\degree$C and 3.4 J/cm$^2$. Since our deposition temperatures and availability of atomic species place the process near a region of the phase diagram where $\beta$-FeSe and Fe$_7$Se$_8$ are adjacent to each other, mixed configurations of these two phases are likely, especially given the stoichiometry fluctuations expected in the PLD flux. Kinetic effects during crystal growth are also certain to be significant under the nonequilibrium conditions of PLD. The co-existence of $\beta$-FeSe and Fe$_7$Se$_8$ has also been well documented in bulk crystals \cite{Williams2009}.\par

The film grown with laser energy density 1.4 J/cm$^2$ (Fig. \ref{Fig 1.}a) shows two new prominent reflections with respect to the film grown at 3.4 J/cm$^2$, that cannot be attributed to the $\beta$ phase. They appear at 2$\theta$ = 32.17$\degree$ ($d$ = 2.780 $\AA$) and 2$\theta$ = 67.30$\degree$ ($d$ = 1.390 $\AA$). The 2:1 ratio of their interplanar separations manifested in this symmetric scan indicates that the two peaks are due to a common oriented phase. Because of their shared NiAs-type fundamental structure, $\delta$-FeSe and Fe$_7$Se$_8$ both feature strong reflections at these positions. Our relatively low growth temperature makes the formation of the high-temperature stoichiometric $\delta$ phase unlikely \cite{Okamoto, McQueen2009ex,Jain1978}. Fe$_7$Se$_8$, on the other hand, is known to be stable in our temperature range, has already been observed in the 3.4 J/cm$^{2}$ film, and is consistent with the Se-enriched composition of our PLD target. Accordingly, we assign the two new prominent peaks seen in the $\theta$-2$\theta$ scan to the (101) and (202) reflections of the hexagonal NiAs-type fundamental structure of Fe$_7$Se$_8$. Corroborating evidence of the (101) orientation of this Fe$_7$Se$_8$ phase comes from thin film 2$\theta$ scans (Supplementary material). 2$\theta$ scans ($\omega$ = 2$\degree$) of the 1.4 J/cm$^{2}$ film reveal a reflection at 2$\theta$ = 42.5$\degree$, corresponding to a lattice plane with inclination of 19.2$\degree$ with respect to the surface normal. This measured peak matches the (102) reflection of the fundamental NiAs-type structure of Fe$_7$Se$_8$ and confirms the epitaxial orientation of the Fe$_7$Se$_8$ phase along the (101) direction of the fundamental structure. This (102) reflection was used to generate the $\phi$ scan labeled as Fe$_7$Se$_8$ (102) in Fig. \ref{fig:my_label}. As emphasized in the above discussion of the reflections labeled as (101) and (202) in Fig. \ref{Fig 1.}a, all Miller indices used in this paper to specify the directions of the Fe$_7$Se$_8$ crystal, refer to the setting of its fundamental NiAs-type structure, unless otherwise noted. The $\beta$-FeSe (001) and (003) reflections are still clearly present in the 1.4 J/cm$^{2}$ film, whereas the (002) and (004) reflections are now convoluted with the reflections of Fe$_7$Se$_8$. This film is a mixture of hexagonal Fe$_7$Se$_8$ and tetragonal $\beta$-FeSe, both of which are epitaxially oriented. The Fe$_7$Se$_8$ phase grows with the relation Fe$_7$Se$_8$ (101)$\parallel$(001) MgO. A representative $\phi$ scan of the Fe$_7$Se$_8$ (102) reflection shown in Fig. \ref{fig:my_label} indicates four-fold symmetry with two peaks of distinct intensities alternating every 45$\degree$. This defines two separate epitaxial domains: one aligned with the cubic unit cell of the substrate and one rotated by 45$\degree$. We conclude the in-plane orientation of these domains may be expressed as Fe$_7$Se$_8$ [010]$\parallel$[100] MgO and Fe$_7$Se$_8$ [010]$\parallel$[110] MgO. The in-plane orientation behavior noted in Fig. \ref{fig:my_label} for Fe$_7$Se$_8$ is also observed in all other films in this paper that contain this phase. The pattern of the 1.4 J/cm$^2$ film in Fig. \ref{Fig 1.}a shows an additional peak near 33.7$\degree$, which is marked by a dagger ($\dagger$). This peak is also consistent with a reflection of Fe$_7$Se$_8$ and suggests the presence of a minor fraction of this phase in yet another orientation. Because Fe$_7$Se$_8$ has several possible ordered vacancy superstructures, leading to numerous diffraction peaks, and is epitaxially oriented in our films, it is difficult to accurately assign Miller indices to all reflections of this phase. It should be noted however, that none of the peaks here and subsequently assigned to Fe$_7$Se$_8$ can be ascribed to any reasonable contaminant phases such as iron oxide, iron, or selenium.\par

Figure \ref{Fig 1.}b shows $\theta$-2$\theta$ XRD scans for a series of thin films grown using a laser energy density of 3.4 J/cm$^{2}$ with various substrate temperatures. Films grown at 350-450$\degree$C show again the concomitant presence of $\beta$-FeSe and Fe$_7$Se$_8$. Starting at 350$\degree$C and increasing to 450$\degree$C, the relative fractions of $\beta$-FeSe and Fe$_7$Se$_8$ vary, with $\beta$-FeSe preferred at lower temperatures and Fe$_7$Se$_8$ favored at higher temperatures. The $\beta$ phase is (001)-oriented while Fe$_7$Se$_8$ grows with the (101) orientation of its fundamental NiAs-type structure. The corresponding $\phi$ scans for the $\beta$ (103) and Fe$_7$Se$_8$ (102) reflections are identical to those shown in Fig. \ref{fig:my_label}, establishing the double epitaxy of tetragonal and hexagonal phases in these films. At 500$\degree$C, the previous double epitaxy trend is broken and the $\beta$ phase is formed  without the (101) orientation of the Fe$_7$Se$_8$ phase. Increasing the temperature to 550$\degree$C causes a change in the orientation of the $\beta$-FeSe epilayer from $\beta$-FeSe (001)$\parallel$(001) MgO to $\beta$-FeSe (101)$\parallel$(001) MgO. We note that all films shown in Fig. \ref{Fig 1.}b also contain minor fractions of Fe$_7$Se$_8$ in the $c$-axis orientation previously discussed in the context of Fig. \ref{Fig 1.}a, and marked by asterisks in both figures and throughout the paper.

\begin{figure}[t]
    \includegraphics[width=7.8cm]{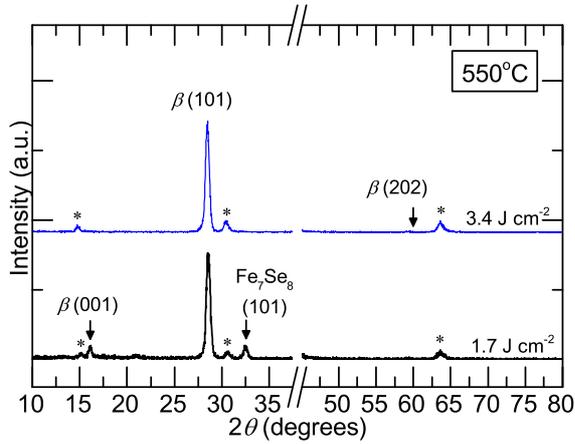}
    \caption{$\theta$-2$\theta$ XRD patterns of FeSe films grown at a substrate temperature of 550$\degree$C show that the dominant (101) $\beta$-FeSe orientation is maintained at the lower laser fluence of 1.7 J/cm$^2$, although a greater phase heterogeneity is noted with additional (001)-oriented $\beta$-FeSe and (101)-oriented Fe$_7$Se$_8$ content detected. Reflections marked with asterisks ($\ast$) indicate the same Fe$_7$Se$_8$ reflections noted in previous figures.}
    \label{Fig 3.}
\end{figure}
\begin{figure}
    \includegraphics[width=7.8cm]{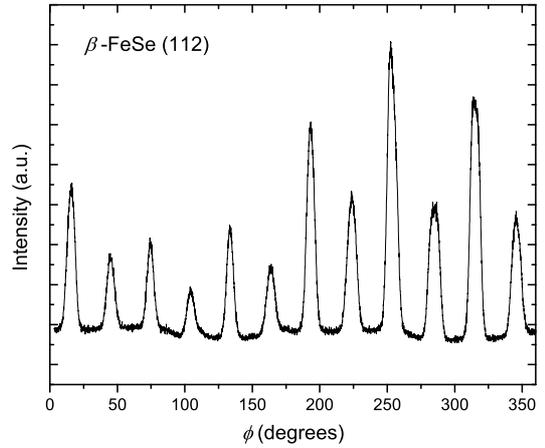}
    \caption{$\phi$ scan of the noncoplanar (112) reflection on the (101)-oriented $\beta$-FeSe film shows twelve poles which indicates the growth of epitaxial domains with three different in-plane orientations. }
    \label{112scan}
\end{figure}
\begin{figure}
    \includegraphics[width=7.8 cm]{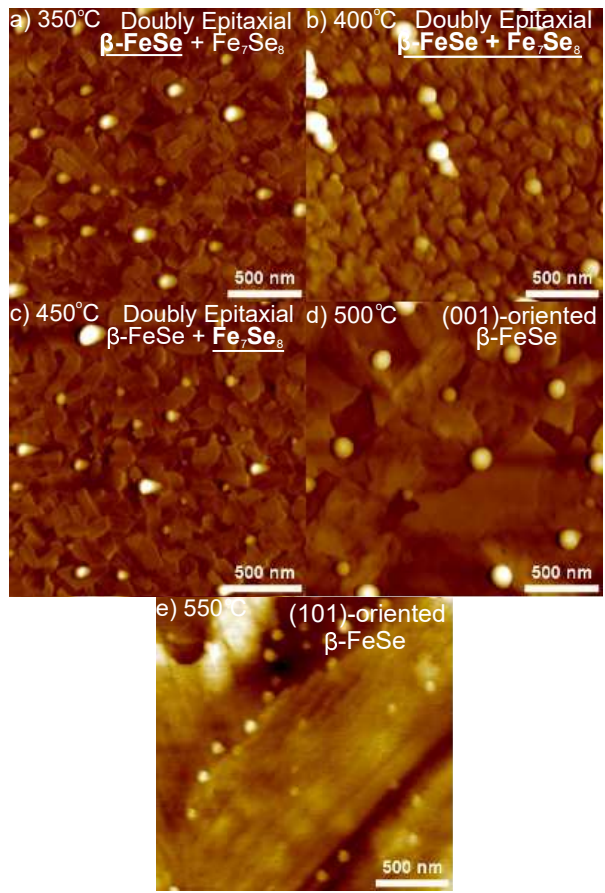}
    \caption{Atomic force microscopy images of FeSe films grown with a laser fluence of 3.4 J/cm$^2$ at various substrate temperatures. Corresponding XRD patterns are shown in Fig. \ref{Fig 1.}b for respective temperatures. The doubly epitaxial films containing $\beta$-FeSe and Fe$_7$Se$_8$ that are grown at a) 350$\degree$C, b) 400$\degree$C, and c) 450$\degree$C show a fine granular surface consistent with interspersed crystal domains of the $\beta$ and Fe$_7$Se$_8$ phases. Corresponding XRD patterns indicate that these domains have well-defined epitaxial relationships to each other and to the substrate. Predominantly (001)-oriented $\beta$-FeSe films without the presence of the (101)-oriented Fe$_7$Se$_8$ are obtained at d) 500 $\degree$C and show larger domains with a smoother surface. (101)-oriented $\beta$-FeSe films grown at e) 550 $\degree$C have significantly larger striated grains.}
    \label{Fig 5.}
\end{figure}

\begin{figure}
    \includegraphics[width = 7.8cm]{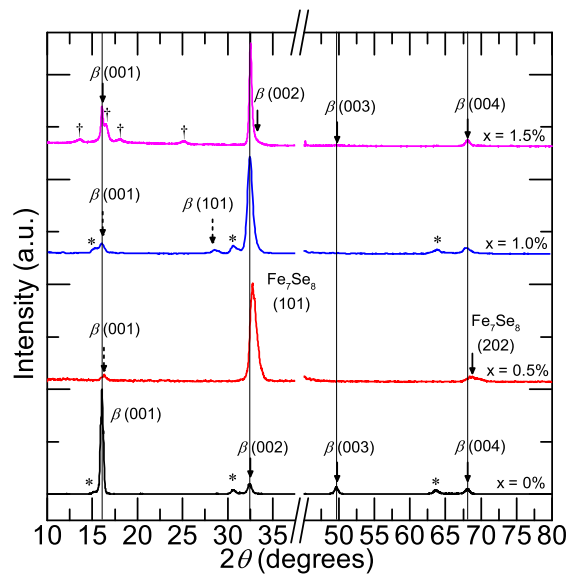}
    \caption{$\theta$-2$\theta$ scans of thin films with nominal Fe$_{1-x}$Cu$_x$Se composition. Films grown with Cu doping via the addition of Cu$_2$Se powder to the PLD targets are dominated by Fe$_7$Se$_8$. $\beta$-FeSe appears to be present in small amounts in the films with 0.5 at.\% and 1.0 at.\% Cu. At 1.5 at.\% Cu doping, (001)-oriented $\beta$-FeSe and (101)-oriented Fe$_7$Se$_8$ are observed in a doubly epitaxial configuration. Peaks marked with daggers ($\dagger$) are consistent with other Fe$_7$Se$_8$ reflections.}
    \label{Fig 4.}
\end{figure}
\begin{figure}[t]
    \includegraphics[width=7.8 cm]{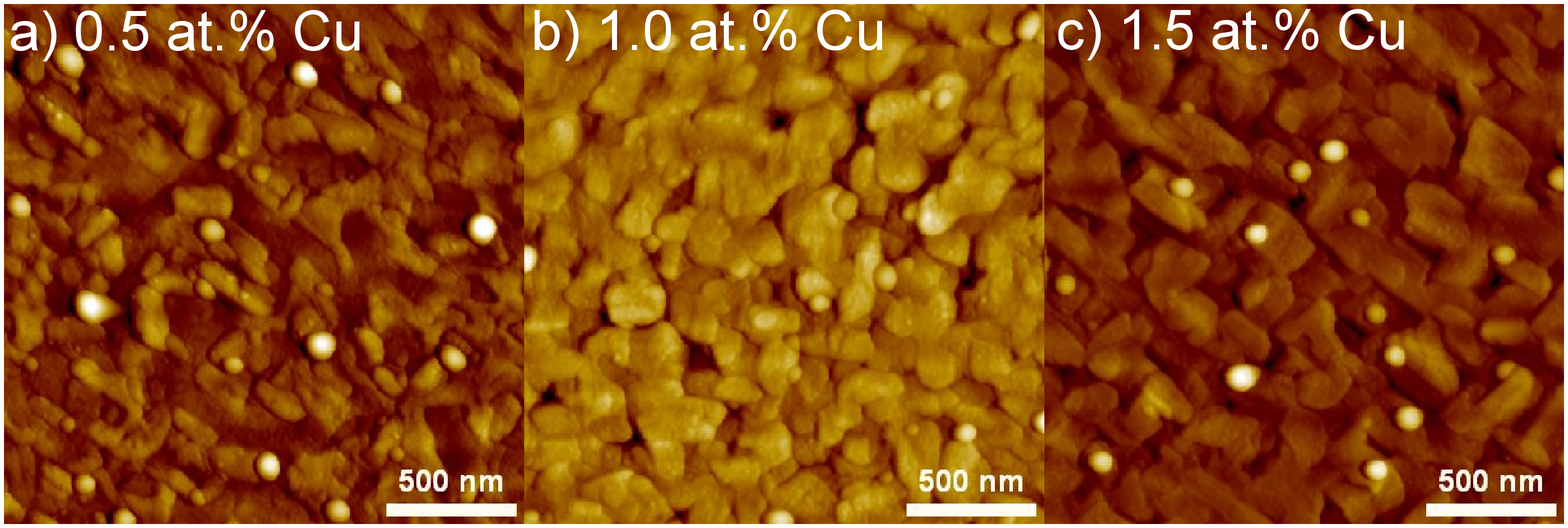}
    \caption{Atomic force microscopy images of films with nominal Fe$_{1-x}$Cu$_x$Se composition for three different Cu concentrations grown with a substrate temperature of 500$\degree$C and laser fluence of 3.4 J/cm$^2$. The films grown with a) $x = 0.5$ at.\% Cu and b) $x = 1.0$ at.\% Cu are predominantly (101)-oriented Fe$_7$Se$_8$. Their surfaces are rougher relative to undoped films. Although majority Fe$_7$Se$_8$, the film doped with c) $x = 1.5$ at.\% Cu is doubly epitaxial comprising (001)-oriented $\beta$-FeSe and (101)-oriented Fe$_7$Se$_8$}
    \label{Fig 7.}
    \end{figure}
The change in growth behavior around 500$\degree$C suggests there are competing growth processes taking place that are sensitive to temperature between 450$\degree$C and 550$\degree$C.
This change in crystallographic orientation appears to indicate the existence of a growth regime that has not been fully explored before.
In order to further evaluate the shift in orientation from (001) to (101) at 550$\degree$C, more films were grown at 550$\degree$C with a lower laser fluence. 
As shown in Fig. \ref{Fig 3.}, the lower laser fluence of 1.7 J/cm$^2$ also yields a (101)-oriented $\beta$-FeSe thin film when growth takes place at 550$\degree$C. 
The lower laser fluence leads to greater phase heterogeneity with peaks ascribed to (001)-oriented $\beta$-FeSe and (101)-oriented Fe$_7$Se$_8$ appearing in addition to the phases present at 3.4 J/cm$^2$. 
To determine the in-plane orientation, a noncoplanar $\phi$ scan of the (112) reflection was carried out with inclination $\chi = 35\degree$ and is shown in Fig. \ref{112scan}. Since the (112) plane has 4-fold rotational symmetry, the 12 poles observed indicate 3 epitaxial domains, one aligned with the substrate and the other two rotated $\pm$ 30$\degree$ with respect to the $a$-axis of the substrate. Similar to the in-plane relationship with the (001) orientation, there exist several possibilities for a configuration with a mismatch that is acceptable for epitaxial growth. The spacing between iron atoms in the (101) plane is 6.624 \AA. Consequently, 7 unit cells of (101)-oriented $\beta$-FeSe matches to 11 unit cells of MgO with an overall mismatch of 0.10\%, as shown in Fig. \ref{Fig 2.}b, when aligned with the $a$-axis of MgO.

The surface morphology of the FeSe films was analyzed by AFM. AFM scans of the films whose diffractograms are shown in Fig. \ref{Fig 1.}b can be seen in Fig. \ref{Fig 5.}. The lower temperature depositions show a granular morphology that is consistent with the double epitaxy of $\beta$-FeSe and Fe$_7$Se$_8$. Although the ``grains'' defined by the height variations in the AFM images are unlikely to represent the actual crystal grains of the film, the height variation ``grain structure'' in AFM is a manifestation of the expected underlying configuration of interspersed crystal domains of $\beta$-FeSe and Fe$_7$Se$_8$ with well-defined epitaxial relationships to each other and to the substrate. Surface roughness is 4.05 nm and 3.73 nm for 350$\degree$C and 400$\degree$C, respectively. The film grown at 450$\degree$C (Fig. \ref{Fig 5.}c) has smaller step-heights across the ``grains'' and the film roughness decreases relative to the lower temperature films to 2.97 nm. Referring to Fig. \ref{Fig 1.}b, the films grown at 350-450$\degree$C are changing from majority (001)-oriented $\beta$-FeSe (Fig. \ref{Fig 5.}a) to majority (101)-oriented Fe$_7$Se$_8$ (Fig. \ref{Fig 5.}c) as the temperature increases. The morphology of the (001)-oriented $\beta$-FeSe film grown at 500$\degree$C shows an increase in ``grain'' size. In this case, where (101)-oriented Fe$_7$Se$_8$ is no longer present, the ``grain structure'' exhibited in the AFM scan is likely due to the slight misorientation of the individual $\beta$-FeSe grains that lead to the mosaicity of the film. It is noteworthy that the roughness of the 500$\degree$C film increases to 7.15 nm. The higher roughness value of this film is determined by the large ``spherical-like'' objects on the surface (also evidenced in all other AFM scans). The regions in between the ``spheres'' in the 500$\degree$C film are smoother than the lower-temperature samples. Finally, at 550$\degree$C, the crystal structure changes to (101)-oriented $\beta$-FeSe and this is reflected in the morphology showing substantially larger, striated ``grains.''\par

The addition of dopants such as Cu in the FeSe system, which is important to modulate the electrical properties, may affect crystal growth.
To explore the effects of chemical doping on double epitaxy behavior, Cu-doped thin films with nominal composition Fe$_{1-x}$Cu$_x$Se were grown at a substrate temperature of 500$\degree$C with a laser fluence of 3.4 J/cm$^2$. The XRD patterns in Fig. \ref{Fig 4.} show that the film doped with nominal Cu concentration $x=0.5$ at.\% exhibits the (101)-oriented Fe$_7$Se$_8$ phase with the (101) and (202) reflections of the fundamental NiAs-type structure near 32.2$\degree$ and 67.3$\degree$, respectively. A very small amount of $\beta$-FeSe also seems to be present in the $x=0.5$ at.\% Cu film. Fe$_7$Se$_8$ remains the dominant phase upon intended further doping up to the maximum nominal concentration explored, which was 1.5 at.\% Cu. The orientation of all Cu-doped films is consistent with the same domain-matched epitaxial growth as the undoped samples with the relationship Fe$_7$Se$_8$ (101)$\parallel$(001) MgO.  The dominance of the Fe$_7$Se$_8$ phase is likely due to an altered Fe:Se ratio in the PLD flux produced from the doped targets. Further studies using PLD plasma plume diagnostics and compositional analysis of the targets are needed to determine whether this effect is due to the presence of Cu, its Cu$_2$Se precursor, or other nonstoichiometric factors in the doped targets. Other aspects of the growth process may also be involved including, for example, changes in laser-target interaction due to Cu addition (e.g., variation in target surface roughness from Cu-induced changes in target sintering behavior), kinetic effects during film growth due to the presence of Cu, and changes in $\beta$ phase stability under PLD conditions when Cu is present. It is noteworthy, however, that the increase in Cu concentration seems to correlate with the emergence of the $\beta$ phase in the doped films. At $x = 1.5$ at.\% Cu, (001)-oriented $\beta$-FeSe and (101)-oriented Fe$_7$Se$_8$ are once again observed in a doubly epitaxial configuration. Judging from the interplay between $\beta$-FeSe and Fe$_7$Se$_8$ seen in the undoped films, there is likely a combination of substrate temperature and laser fluence that could favor the dominance of the $\beta$-FeSe phase also in the doped films. \par
AFM scans of the Cu-doped films are shown in Fig. \ref{Fig 7.}. Cu doping leads to films with greater surface roughness (7.0-7.3 nm range) than undoped films. These films show a granular structure, comparable to the doubly-epitaxial undoped films in Fig. \ref{Fig 5.}a-c. The grain shape is different than previously seen, with an elongated shape and preferred direction of the longer axis. Once again ``spherical-like'' objects are also noted on the surface, suggestive of a concomitant process of island nucleation and growth that merits further investigation.

\section{Conclusion}
\label{S:4}
We have studied the conditions which result in double epitaxial growth of tetragonal and hexagonal phases of FeSe on MgO during PLD. The nonstoichiometry of Fe:Se in the PLD flux, allied to the existence of numerous domain matching relationships between the FeSe phases and MgO, provide favorable conditions for the concomitant growth of multiple phases of varied crystallographic orientations. These conditions can be accessed by selecting suitable combinations of laser fluence and substrate temperature. We have shown that the simultaneous, epitaxial growth of (001)-oriented $\beta$-FeSe and (101)-oriented Fe$_7$Se$_8$ occurs for substrate temperatures in the 350-450$\degree$C range for ablation of a Se-rich target under a laser fluence of 3.4 J/cm$^2$. In this double epitaxy, the $\beta$-FeSe phase grows with the base of its tetragonal unit cell aligned with the cubic substrate, while the Fe$_7$Se$_8$ domains take on two different in-plane orientations, which are rotated from each other by 45$\degree$. Higher substrate temperatures of 500-550$\degree$C suppress (101)-oriented Fe$_7$Se$_8$ and allow $\beta$-FeSe films to be selected between (001)-oriented and (101)-oriented with respect to the substrate. Altering the chemical composition of the FeSe PLD targets by the addition of Cu$_2$Se powder results in films dominated by the Fe$_7$Se$_8$ phase for doping levels of 0.5-1.0 at.\% Cu. Little $\beta$ phase is observed in the films in this low-doping range. For a Cu concentration of 1.5 at.\% however, $\beta$-FeSe reemerges in the films, with the doubly epitaxial configuration of the $\beta$ and Fe$_7$Se$_8$ phases observed again. Doubly epitaxial films show surface morphology with granular structure, consistent with interspersed crystal domains of $\beta$-FeSe and Fe$_7$Se$_8$. Thin films dominated by $\beta$-FeSe in which double epitaxy is suppressed are generally smoother with larger grains, consistent with their mosaicity indicated by XRD. ``Spherical-like'' objects are observed on the surface of all films. These structures are suggestive of island nucleation and growth processes. The number density and size of these islands may be controllable by further adjustments in the laser fluence incident on the PLD targets. Further surface characterization of the doubly epitaxial films using magnetic force microscopy (MFM) may allow observation of crystal phase boundaries based on the different magnetic responses of $\beta$-FeSe and Fe$_7$Se$_8$. MFM may also help clarify the nature of the observed spherical-like features. Finally, PLD plasma plume diagnostics may shed light on how to control the content, kinetic energy, and ionization state of Fe and Se in the PLD flux. Such control should enable a deeper understanding of the film growth kinetics during double epitaxy of FeSe and other similar systems.

\section*{Acknowledgments}
The authors wish to thank David J. Hilton, Ashlyn D. Burch, Sergey B. Mirov, and Vladimir V. Fedorov for helpful discussions. We are also thankful to S. Aaron Catledge and Tracy G. Hastings for assistance with the AFM measurements. SBH acknowledges graduate fellowship support from the NASA Alabama Space Grant Consortium (ASGC) under award NNX15AJ18H. This work was supported in part by the Air Force Office of Scientific Research (AFOSR) under award FA9550-13-1-0234 and by the NSF Major Research Instrumentation (MRI) Grant No. DMR-1725016. Any opinions, findings, and conclusions or recommendations expressed in this material are those of the authors and do not necessarily reflect the views of the National Science Foundation or the Air Force Office of Scientific Research.

\bibliographystyle{apsrev}
\bibliography{main.bib}

\end{document}